\documentstyle[11pt]{article}
\begin{document}

\centerline{\bf\Large Study of the $B_c$-meson lifetime}

\vspace{1cm}

\centerline{\small Chao-Hsi Chang$^{1,2}$, Shao-Long Chen$^3$,
Tai-Fu Feng$^{2,3}$ and Xue-Qian Li$^{1,2,3}$}

\vspace{0.5cm}

\begin{center}
{\small 1. CCAST (World Laboratory), P.O. Box 8730, Beijing 100080,
China\footnote{Not postmail address for the authors.}}
\\
{\small 2. Institute of Theoretical Physics, Academy of Sciences,
Beijing 100080, China}
\\
{\small 3. Department of Physics, Nankai University, Tianjin 300071,
China}
\end{center}

\vspace{2cm}

\begin{center}
\begin{minipage}{12cm}

\noindent{\bf Abstract}

\vspace{0.2cm}

We in terms of optical theorem estimate the lifetime of $B_c-$meson with
the parameters which are deturmined by fitting the data for the lifetimes
and inclusive semilepton-decays of various $B$ and $D$ mesons. In the
estimate, we find that the bound-state effects are important, and take
them into account carefully in the framework which attributes the effects
to the effective masses of the decay heavy quarks in the inclusive
processes. We also find that to $B_c$ lifetime the penguin contribution is
enhanced due to possible interference between the penguin and the `tree
part' $c_1O_1+c_2O_2$.


\end{minipage}
\end{center}

\vspace{2cm}

\baselineskip 22pt

Very recently the meson $B_c$ has been observed
in CDF detector at Fermilab Tevatron. The observation is
through the semi-leptonic decays $B_{c}{\longrightarrow}
{J/\psi}+ l +\nu_{l}$, and not only the value
of its mass $m_{B_c} = 6.40\pm 0.39 \pm 0.13$ GeV, but also the lifetime
$\tau_{B_c} = 0.46^{+0.18}_{-0.16} \pm 0.03$ ps are given \cite{cdf}.
Therefore to estimate its lifetime so as to understand the meson and
its decay mechanisms becomes one re-freshed interesting problem.

$B_c-$meson is composed of two heavy flavors and both of them
contribute to the lifetime comparetively. It is known that
the Heavy Quark Effective Theory (HQET) \cite{Wise} successfully applies
to phenomenology of heavy mesons and baryons, although there are still some
open problems in B and D physics, such as the unexpected difference
between lifetimes of $\Lambda_b$ and B-mesons \cite{Al} and missing charm
puzzle \cite{Lenz} etc. Since there is no light flavor quarks
in $B_c-$meson, HQET so the Isgur-Wise form factor scenario does
not apply here.

Since the optical theorem may apply to inclusive processes
properly so some non-perturbative effects can be absorbed by
the theorem, thus we focus our attention on the inclusive
processes, especially, the lifetime with the help of the
theorem in this paper.

As realized by many authors, in D and B decays besides the
spectator mechanism i.e. the direct decay of $c$ or $\bar b$, the
non-spectator mechanisms are quite important too. There are two
distinct types of non-spectator mechanisms: the W-annihilation
(WA) and the Pauli-interference (PI), whose details are depicted
in \cite{Neu}. As for $B_c$ meson, owing to the non-spectator
mechanisms the situation becomes more complicated and interesting.
In $B_{d(u)}, B_s$ decays, the penguin contributions and its
interference with that of the `tree piece' $c_1O_1+c_2O_2$ can be
ignored due to the CKM suppression. In contrary, for $B_c$ meson,
$\bar{b} \rightarrow \bar{c} +``W^{+}"$ and $``W^{+}" \rightarrow
c + \bar{s}$ (here $``W^{+}"$ denotes a virtual $W^+-$boson) are
favored according to the CKM entries, so there are charm in
initial state and charm, anti-charm in final state. Therefore, the
interference between the penguin and `tree piece' may contributes
to the decay (see below for some details). Namely the penguin
contribution becomes more important and $B_c-$meson decay may
serve as an ideal place to study the penguin mecahnism. Our
numerical results show that such interference can cause a
contribution as large as about 3$\sim$4\% in the total width,
while in $B_{d(u)}$ and $B_s$ cases, the contribution is less than
0.5\%.

Moreover, in $B_c$, as in the heavy quarkonia, the bound-state effects
should be considered carefully, even when evaluating its inclusive
processes. We take a phenomenological approach to deal
with the masses of the decay quarks $b (\bar b)$ and $c (\bar c)$ in
various heavy mesons i.e. in various bound-state by fitting
data\cite{Chang}. Recently, a modification of HQET,
the so-called Heavy Quark Effective Field Theory (HQEFT), is
proposed\cite{Wu}, where the bound state effects are taken into
account for $B-$ hadrons:
when evaluating the decays, the b-quark mass takes a
different value in $\Lambda_b$ from that in B-mesons, so the
aforementioned problem, the lifetime difference between $\Lambda_b$
and $B-$mesons, can be answered reasonably.
Indeed, the bound state effects, which affect
the inclusive decays, may be attributed to the effective mass of
the decay quark mainly.

For $B_{d(u)}, B_s$ decays, the contributions to the decay widths
from $\bar b$ and $c$ quarks are described quite well in
literature\cite{Bigi,Bigi1,Altarelli}, and the general formulation
for the non-spectator contributions WA and PI
have been given in \cite{Neu}. The formulation of $B_c$ is given
in \cite{Chang} and the readers who are interested
in the details are advised to refer to it. Here we only present
the part of formulation relating to $B_c$ for later convenience,
focus on description of the physical picture,
and discuss the physical essence and consequences.

The spectator contributions to $B_c$ lifetime are the incoherent
sum of $\bar b$ and $c$ decay, when the bound-state effects,
as pointed out above, are attributed to the effective masses
of $\bar b$ and $c$ quarks respectively only,
that is one of essential differences from the other $B-$mesons.
\begin{equation}
\Gamma^{\rm spectator}=\Gamma_b^{\rm spectator}+\Gamma_c^{\rm
spectator}.
\end{equation}
Furthermore, the non-spectator parts may have a unignorable interference.

The effective Lagrangian, on that the present study is based, is
\begin{eqnarray}
      \nonumber
 L_{eff}^{\Delta B=1} &=& -\frac{4G_F}{\sqrt2} {\Big \{}
   V_{cb}[V_{ud}^*(c_1(\mu)O_1^u+c_2(\mu)O_2^u)+
           V_{cs}^*(c_1(\mu)O_1^c+c_2(\mu)O_2^c)+
  \\
  & &
  \sum\limits_{l=e,\tau,\mu}\bar l\gamma_\mu L\nu\bar c\gamma^\nu L
  b+V_{cs}^\ast\sum\limits_{i=3}^6 c_iO_i]
  {\Big \}}+h.c.
\end{eqnarray}
Here the notations for operators and their coefficients
are given in \cite{He}.

Skipping the tedious details, we have the contribution of
PI i.e. $\Gamma^{PI}(tree)$ and $\Gamma^{PI}(penguin)$ to the total width as
\begin{eqnarray}
{\bf \Gamma}^{PI}(tree)&=&
         \frac{G_F^2}{4\pi}f_{Bc}^2M_{B_c}|V_{cb}|^2|V_{cs}|^2(1-z_-)^2p_-^2
            \cdot  {\Big \{}
           [2c_1c_2+\frac{1}{N}(c_1^2+c_2^2)]B_1
        \nonumber \\
       && +2(c_1^2+c_2^2)\epsilon_1
         {\Big \}}
    \\
{\bf\Gamma}^{PI}(penguin)&=&
        \frac{G_F^2}{4\pi}f_{Bc}^2M_{B_c}|V_{cb}|^2|V_{cs}|^2(1-z_-)^2p_-^2
          \cdot        {\Big \{}
            [2c_2c_4+2c_1c_3+2c_3c_4+
       \\    \nonumber
       & &
             \ \    \frac{1}{N}(c_3^2+c_4^2+2c_2c_3+2c_1c_4)]B_1
                 +2(c_3^2+c_4^2+2c_2c_3+2c_1c_4)\epsilon_1
        {\Big \}}
   \\     \nonumber
   & &
    -\frac{G_F^2}{4\pi}f_{Bc}^2M_{B_c}|V_{cb}|^2|V_{cs}|^2(1-z_-)^2
          \cdot     {\Big \{}
          [2c_5c_6+\frac{1}{N}(c_5^2+c_6^2)][\frac{2+z_-}{3}p_-^2\tilde B_2
   \\      \nonumber
   & &
        -\frac{1+2z_-}{6}(m_b^2\tilde B_1+m_c^2B_1-4m_bm_cB_2+2m_bm_cB_1)]
           +2(c_5^2+c_6^2)[\frac{2+z_-}{3}p_-^2\tilde\epsilon_2
   \\        \nonumber
   & &
             -\frac{1+2z_-}{6}(m_b^2\tilde\epsilon_1
             +m_c^2\epsilon_1-2m_bm_c(\epsilon_3+\epsilon_4)
              +m_bm_c(\epsilon_5+\epsilon_6))]
           {\Big \}}
    \\    \nonumber
   & &
          -\frac{G_F^2}{8\pi}f^2_{Bc}M_{B_c}|V_{cb}|^2|V_{cs}|^2(1-z_-)^2
          \bar m_c \cdot    {\Big \{}
           [c_1c_5+c_2c_6+c_3c_6+c_4c_5
     \\   \nonumber
     & &
           +\frac{1}{N}(c_1c_6+c_2c_5+c_4c_6+c_3c_5)][2m_cB_1
           +m_b(-4B_2+2B_1)]
       \\
      & &
           +2(c_1c_6+c_2c_5+c_4c_6+c_3c_5)[2m_c\epsilon_1-2m_b(\epsilon_3
           +\epsilon_4) +m_b(\epsilon_5+\epsilon_6)]
             {\Big \}}.
\end{eqnarray}
For WA, we have three pieces as the follows:
\begin{eqnarray}
        \nonumber
   {\bf\Gamma}^{WA}(tree)&=&
        -\frac{G_F^2}{12\pi}|V_{cb}|^2|V_{cs}|^2f_{Bc}^2M_{B_c}(1-z_+)^2
        \cdot      {\Big \{}
              [Nc_1^2+2c_1c_2+\frac{c_2^2}{N}]
     \\   \nonumber
       & &
                \times [(1+\frac{z_+}{2})M_{B_c}^2 B_1
                      -(1+2z_+)(m_b^2B_2+m_c^2\tilde B_2+2m_bm_c B_2)]
       \\
       & &
            +2c_2^2[(1+\frac{z_+}{2})M_{B_c}^2 \epsilon_1
                     -(1+2z_+)(m_b^2\epsilon_2+m_c^2\tilde \epsilon_2
                     +m_bm_c(\epsilon_3+\epsilon_4))]
           {\Big \}},
      \\        \nonumber
{\bf\Gamma}^{WA}(B_c\rightarrow \tau\nu)&=&
          -\frac{G_F^2}{12\pi}|V_{cb}|^2|V_{cs}|^2f_{Bc}^2M_{B_c}(1-z_+)^2
          \cdot      {\Big \{}
               (1+\frac{z_\tau}{2})M_{B_c}^2 B_1
       \\
       & &
               -(1+2z_\tau)(m_b^2B_2+m_c^2\tilde B_2+2m_bm_c B_2)
           {\Big \}},
\end{eqnarray}
\begin{eqnarray}
       \nonumber
{\bf\Gamma}^{WA}(penguin)
         &=&
           -\frac{G_F^2}{12\pi}|V_{cb}|^2|V_{cs}|^2f_{B_c}^2M_{B_c}(1-z_+)^2
          \cdot    {\Big \{}
                [(\frac{2c_2+c_3}{N}+2c_1+c_4)(c_3+Nc_4)]
           \\       \nonumber
           & &
                \times [(1+\frac{z_+}{2})M_{B_c}^2 B_1
                 -(1+2z_+)(m_b^2B_2+m_c^2\tilde B_2+2m_bm_c B_2)]
       \\       \nonumber
       & &
             +2(2c_2+c_3)c_3\cdot [(1+\frac{z_+}{2})p_+^2\epsilon_1
                  -(1+2z_+)(m_b^2\epsilon_2+m_c^2\tilde \epsilon_2
                  +m_bm_c(\epsilon_3+\epsilon_4))]
          {\Big \}}
       \\        \nonumber
        & &
           +\frac{G_F^2}{2\pi}|V_{cb}|^2|V_{cs}|^2f_{Bc}^2
           M_{B_c}^3(1-z_+)^2 \cdot    {\Big \{}
                   [\frac{c_5^2}{N}+2c_5c_6+Nc_6^2]\tilde B_2
                   +2c_5^2\tilde\epsilon_2
           {\Big \}}
       \\     \nonumber
       & &
           -\frac{G_F^2}{4\pi}|V_{cb}|^2|V_{cs}|^2f_{Bc}^2
           M_{B_c}\bar m_c(1-z_+)^2 \cdot     {\Big \{}
               [(\frac{c_2+c_3}{N}+c_1+c_4)(c_5+Nc_6)]
      \\
      & &
             \times [2m_b B_2+2m_c\tilde B_2]
             +2(c_2+c_3)c_5\cdot [m_b(\epsilon_3+\epsilon_4)+2m_b
             \tilde\epsilon_2]
            {\Big \}},
\end{eqnarray}
where
\begin{eqnarray}
      z_-=\frac{\bar m_c^2}{p_-^2}, &&
      p_-=p_b-p_{\bar c},
\end{eqnarray}
and
\begin{eqnarray}
      \nonumber
    p_+ &=& p_b+p_c,
   \\   \nonumber
    z_+ &=& \frac{\bar m_c^2}{p_+^2}=\frac{\bar m_c^2}{M_{B_c}^2},
    \\
    z_\tau &=& \frac{m_\tau^2}{p_+^2}=\frac{m_\tau^2}{M_{B_c}^2}.
\end{eqnarray}

Note that due to the chiral suppression only $B_c\to \tau^+\nu$ and
$B_c\to c\bar s$ are considered for WA.

In the expressions the pieces with the superscripts `WA' or `PI'
(corresponding to the W-annihilation or Pauli-interference parts)
and {\it tree} in the brackets denote all the contributions from the
`tree' part, i.e. the $c_1O_1$ and $c_2O_2$ terms, whereas those with
{\it penguin} in the brackets
mean the contributions not only from the $penguin$ part $c_iO_i\;
(i=3,4,5,6)$ alone, but also from its interference with the `tree'
part. Obviously, for $B_c$, there are interferences proportional to
$c_ic_1,c_ic_2\;(i=3...6)$, whereas for $B_{d(u)}, B_s$, proportional
to $c_ic_j\;(i,j=3,...6)$, the interferences among the penguin
terms themselves only. Since
$c_1,c_2$ are of order O(1) and $c_i\;(i=3,...6)$ is of order
O($10^{-2}$), so roughly, $c_ic_j$ is four orders smaller than
$c_1^2, c_2^2, c_1c_2$, but $c_ic_1,c_ic_2$ are only two (even one)
orders smaller, thus the penguin terms may play more significant roles
in $B_c$ decays. Moreover, note that since $m_c$ cannot be neglected
in the derivation of decay-width operators for $B_c$ decays, while the
light flavors are taken as zero-mass fermions for $B_{d(u)}, B_s$ decays,
so several new operators appear.

To calculate the the contributions of the non-spectator parts,
one needs to evaluate the hadronic matrix elements with dimention
$6$ operators even for the lifetime,
which are governed by the non-perturbative QCD, and so far can be
dealt with phenomenologically only. As pointed out by many authors
\cite{Neu}, the non-factorization effects modify the results
of the `vacuum saturation' and the modifications can be described
by introducing in a few parameters $B_k$ (deviating from $1$)
and $\epsilon_k$ (deviating from $0$)\footnote{In fact in Eqs.(3-8)
the modifications from the vacuum saturation have been made
in this way already. The detailed expressions
can be found in \cite{Chang}}. With some symmetry
arguments as in literature, we fix the values for the parameters
so the hadronic matrix elements within tolerable errors.

As aforementioned, Instead of  determining the heavy quark masses
in bound-states from any underlying theory as what \cite{Wu} did,
we phenomenologically take the bound state effects into account of
the quark masses by fitting data of the lifetimes and branching
ratios of inclusive-semileptonic decays for the mesons $B_{d(u)}$
and $B_s$.

We simply write
\begin{equation}
\label{del}
M_Q^{eff}=M_Q^{pole}-\Delta,
\end{equation}
where $\Delta$ manifests the bound-state effects and will be
fixed phenomenologically.

Fitting the data for $B_{d(u)}$ and $B_s$ as well as D-mesons,
we obtain:
$$\Delta^{(D)}_c\equiv m_c^{pole}-m_c^{eff}=0.23\;{\rm GeV};$$
$$\Delta_b^{(B)}\equiv m_b^{pole}-m_b^{eff}=0.11\sim 0.13\;{\rm GeV}.$$
The superscripts (B) and (D) denote that the $\Delta$ for b and
c-quarks are determined by fitting the data of D and B mesons respectively.

For the $B_c$ meson, the spectator contribution is an incoherent sum of
that from individual $\bar b$ and $c$ quark decays while leaving
the other as a spectator. But we should note that when evaluating
this contribution, $m_b$ ad $m_c$ take their effective values at
the energy scales $M_B\sim m_b(m_b)$ for the pole mass $m_b^{pole}(B_c)$
and at $M_D\sim m_c(m_c)$ for the pole mass $m_c^{pole}(B_c)$.
Namely, we determine the effective masses for $\bar b$ and $c$ of the
$B_c-$meson through $B$ and $D$ decays. Whereas the $\bar c$ in the
final state is the decay product of $\bar b-$quark and now we are
considering inclusive processes, so we should take its running mass
for the $\bar c-$quark at the energy scale of $M_B$. For the
non-spectator contributions, i.e. the WA and PI pieces, the
corresponding scale for the anti-charm quark mass of the final state
should be $M_{B_c}$. In the numerical calculations for the present
paper, we take the relevant parameters as follows:
$M_{B_c}=6.25$ GeV, $M_{B_c}^*=6.33$ GeV,
$B_i=\tilde B_i(i=1,2)=1$, $\epsilon_1=\tilde \epsilon_1=-0.14$,
$\epsilon_2=\tilde \epsilon_2=-0.08$, $\epsilon_{3,4}\sim \epsilon_2$,
$\epsilon_{5,6}\sim \epsilon_1$,
and for the decay constant we adopt two possible values
$f_{B_c}=500$ MeV\cite{Quigg} and $f_{B_c}=440$ MeV\cite{Davies}
respectively. For the calculation of PI contribution,
$p_-^2=(p_{\bar b}-p_{c})^2 \sim (2m_b^2+2m_c^2-M_{B_c}^2)$
is adopted. With the parameters, we finally obtain the
numerical results and tabulate them in Table 1.
\begin{table}[ht]
\begin{center}
\begin{tabular}{|c|c|c|c|c|c|c|c|c|}\hline\hline
                  &$\tau_{B_c}$ (ps)              & $\Gamma^{pen.}$
                  & $\Gamma^{b\rightarrow c}$     & $\Gamma^{c\rightarrow s}$
                  & $\Gamma^{WA}$                 & $\Gamma^{PI}$
                  & $\Gamma(\tau\nu)$(ps$^{-1}$)  & $B_{SL}$    \\ \hline
$f_{B_c}=440$MeV &$0.362$                   & $3.4\%$
                 &$22.8\%$                  & $70.9\%$
                 &$13.4\%$                   & $-7.1\%$
                 &$0.078$                   & $8.7\%$  \\  \hline
$f_{B_c}=500$MeV &$0.357$                   &$4.3\%$
                 &$22.4\%$                  &$69.7\%$
                 &$16.9\%$                  &$-9.0\%$
                 &$0.100$                   &$8.4\%$   \\   \hline
\end{tabular}
\end{center}
\caption{This is the result for $B_c$ meson, $\tau_{B_c}$ denotes the
lifetime of $B_c$, $\Gamma^{pen.}$ denotes the enhancement caused by
the penguin contribution. The $\Gamma(\tau\nu)$ denotes the
width of the total leptonic decay, and the $B_{SL}$ denotes the
branching ratio of the semileptonic decay of $B_c$.}
\end{table}
\vspace{0.3cm}

Here the puzzle about the effective masses of $\bar b$ and $c$ emerges
again, because both of them reside in the same bound state $B_c$.
Taking all the above parameters, we would obtain the values presented
in table 1. Now let us consider the bound-state effects on the
effective quark-masses in $B_c$ meson more precisely. Because $B_c$
includes two heavy quarks, the bound-state effects might be different
from those in $B, D$ mesons which contains one light flavor. If we
are tempted to believe that the values $m_c^{eff}$ and $m_b^{eff}$
might be smaller than $m_c^{eff}=1.65$ GeV and $m_b^{eff}=4.9$ GeV
obtained by $B$ and $D$ decays and phenomenologically if $m_c^{eff}(B_c)
=1.55$ GeV, $m_b^{eff}(B_c)=4.85$ GeV instead, we would have
$\tau(B_c)\approx 0.47$ ps, which is closer to the recently measured
$B_c$ lifetime\cite{cdf}. In this case, it will mean
$\Delta^{B_c}_c=0.33$ GeV and $\Delta^{B_c}_b=0.17$ GeV.

The results indicate the importance of the bound-state effects on
the phenomenology. It is easy to understand the fact, i.e.
the rates of direct $\bar b$ and $c$ decays, which dominate the
lifetime of $B_c$ meson, are proportional to $(m_Q^{eff})^5$,
the results are somewhat sensitive to the $m_Q^{eff}-$values.
It is a `good' place to `determine' the quark effective masses
$m_Q$. From the physics picture, the quark effective masses
introduced here should have more precise meaning in the framework
of the potential model and be good to involve certain non-perturbative
QCD effects\cite{chang1}.

As a summary, in this work, we study the effects when decaying
quarks reside in bound states phonomenologically with care.
We collect all the informations gained from heavy mesons $B_{d(u)},
B_s$ and $D$, we further calculate the lifetime of $B_c$ meson.
In the process, we notice that the bound state effects are
important even to the spectator modes. Moreover, because of the
CKM entries, and the interference between the penguin and the
tree piece, the penguin contribution is more important in $B_c$
case than in $B_{d(u)}$ and $B_s$. Therefore,
further experimental progress in measuring $B_c$ lifetime and
branching ratios may provide more informations about the reaction
mechanisms and especially it would make definite hints to the
interesting penguin mechanism etc.\\

\noindent{\bf Acknowledgments:}\\

This work is partially supported by the National Natural Science
Foundation of China.

\end{document}